\documentclass[12pt]{article}
\pdfoutput=1
\usepackage{epsfig}
\usepackage{amsfonts}
\usepackage{amssymb}
\usepackage{xcolor}
\usepackage{amsmath}
\usepackage{upgreek}
\usepackage{graphicx}
\usepackage{cite}
\usepackage{calligra}
\usepackage{caption}
\usepackage{subcaption}
\usepackage{cite}
\usepackage{calligra}
\usepackage{hyperref}
\begin{document}

\begin{center}

{\bf \LARGE Dark energy due to quantum corrections to effective potential} \vspace{1.0cm}

{\bf \large V.A. Filippov$^{1,2}$, R.M. Iakhibbaev$^{1}$, D.I. Kazakov$^{1}$\\[0.3cm] and  D.M. Tolkachev$^{1,3}$}

\vspace{0.5cm}
{\it $^1$Bogoliubov Laboratory of Theoretical Physics, Joint Institute for Nuclear Research, 
  6, Joliot Curie, 141980 Dubna, Russia\\
$^2$ Sarov branch of M.V. Lomonosov Moscow State University, \\ Sarov, Russia \\ 
$^3$ B.I. Stepanov Institute of Physics
National Academy of Sciences of Belarus, \\Minsk, Belarus \\
}
\vspace{0.5cm}

\abstract{In this paper we calculate quantum corrections to the effective potential in different models of inflationary cosmology. We show that quantum corrections lead to a modification of the initial potential uplifting its value at the minimum, which can be interpreted as a cosmological constant/dark energy. We concentrate on the models of $\alpha$-attractors and show that one can naturally get small values of the cosmological constant. In the model of quintessence,  the same mechanism modifies the value  of dark energy at the present time.}\\

\end{center}

\section{Introduction}

Inflationary expansion is one of the most promising models allowing one to describe the properties of the early Universe and its later development  \cite{Guth1980,Starobinsky1980,Martin}. 

 Though the nature of early and late acceleration is still a significant theoretical problem, the simple assumption that the potential energy of a scalar field can be the source of accelerated expansion has been successful in terms of investigation, deriving predictions, and agreement with current observational data \cite{BICEP,PLANCK}. In the simplest realisation of the inflationary scenario, it is generally assumed that after the Universe leaves the inflationary stage and post-inflationary reheating, it remains at minimum potential with a value of $\Lambda \sim 10^{-47}$ GeV$^{4}$ or $10^{-120}$ in Planck units \cite{Akrami:2017cir}. 
One can assume that an acceptable explanation of the current cosmic acceleration is related to the introduction of a field that currently has the value of $\Lambda$ \cite{Copeland:2006wr}. In fact, shortly after the discovery of the current cosmic acceleration at the end of the last century, a class of cosmological models was presented that tries to unify accelerating expansion in early and late times \cite{deHaro:2021swo,Dimopoulos:2017zvq,Brissenden:2023yko}.

A small value of the cosmological constant  $\Lambda $ can be explained by the adoption of the anthropic principle, which removes the question of the meaning of such a small value and refers this question to the problem of landscape in string theory. Then it is enough to have a theoretical possibility of obtaining diverse values of primordial quantum fluctuations in a viable Universe with a small value of the dark energy density selected by the anthropic principle. In this paper, however,  we  follow another point of view and show that quantum corrections to some cosmological models\cite{Kazakov:2022pkc,Kazakov:2023tii} can lead to a significant modification of the behaviour of the initial potential and the appearance of a non-zero ground state energy of the Universe which can be interpreted as a cosmological constant. 

In what follows, we apply the formalism of the effective potential to cosmological models known as $\alpha$-attractors \cite{Kallosh2013,Kallosh:2016gqp, Galante}.
{However, simpler models, say, such as arbitrary degree potentials, we have considered in \cite{Kazakov:2022pkc}. It is also worth noting that we use the effective potential, which allows us to calculate all loop corrections. We consider contributions in the leading logarithmic approximation, which is independent of the subtraction scheme. Hence contributions from leading logarithms are computed unambiguously. Therefore, their consideration even in the non-renormalisable theory is justified.
In this paper, we consider the simplest forms of $\alpha$-attractors which can be represented by the so-called $T$-models with the potential
 \begin{equation} \label{potT}
     V(\varphi)=g \tanh^{n}\left(\frac{\varphi}{\sqrt{6 \alpha} M_{Pl}}\right)=g V_T, 
 \end{equation}
(we will mainly focus on the cases $n=2,4$, and call them the $T^2$, $T^4$ models, respectively) and
$E$-models  with the potential
 \begin{equation} \label{potE}
     V(\varphi)=g \left(1-e^{-\sqrt{\frac{2}{3 \alpha}} \frac{\varphi}{M_{Pl}}}\right)^n=g V_E
 \end{equation}
(we call the cases $n=2,4$ $E^2$- and $E^4$-models, respectively)\cite{Scalisi:2015qga}. Here $\varphi$ is the inflationary scalar field, $M_{Pl}=(8\pi G)^{-1/2}$ is the Planck mass, $g=m_{inf}^2M_{Pl}^2$ is the scale of inflation, where $m_{inf}\approx 10^{-5}M_{Pl}$ is the inflaton mass and $\alpha$ is a positive number.  These models are characterised by an exponentially fast reaching of a plateau at large values of the inflaton field and are successfully used to study the stage of early accelerated expansion of the Universe as well as the dark energy dominated phase \cite{Kallosh:2022feu,Galante,Dimopoulos:2017zvq}. The predictions derived from these models are consistent with the PLANCK and BICEP/Keck \cite{BICEP,PLANCK} observational data. Moreover, introducing a number of e-folding of inflation $N$ (in the slow-roll approximation), one can obtain that  all these models have a point of attraction in the leading approximation in $1/N$ \cite{Kallosh2013,Martin}
\begin{equation}
 r \simeq 12 \alpha/N^2, ~~~~~n_s\simeq 1-2/N,\end{equation}
which for $N=50-60$ and $\alpha \sim 1$ fits the observed values of the slow-roll parameters \cite{PLANCK}. 

Another type of models known as the  quintessence inflation models of $\alpha$-attractors can be represented by the T-model with $n=1$. In this case, the cosmological constant is not really a constant but a  function of the field $\varphi$ at the present time  \cite{Galante,Dimopoulos:2017zvq,Brissenden:2023yko}. 
 
The paper is organized as follows. In the first part, we recall the key ideas for deriving the renormalizable group equation (RG-equation) for effective potentials for non-renormalizable interactions generalizing the approach of \cite{CW}. The RG-equation allows one to sum up all the leading contributions (leading logs) to the effective potential in all orders of perturbation theory. Then we consider the application of this approach to a class of $T$ and $E$ cosmological models  and demonstrate how a non-zero cosmological constant appears as a result of radiative corrections.  We discuss the properties of the the obtained  quantum potentials from the point of view of the inflation scenario.  Finally, we consider a special case of the $T$ model, which is usually used to study the quintessence model.

\section{Effective potential for a general scalar model}

The effective potential $V_{eff}(\varphi)$ is defined as a constant part of the effective action \cite{EA}. It can be seen as supplement for the classical potential by quantum loop corrections. The effective potential can be obtained by summing all 1PI vacuum diagrams with propagators containing a mass term depending on the classical field $\varphi$: $m^2(\varphi)=gv_2(\varphi)$, where $v_2(\varphi)\equiv \frac{d^2V(\varphi)}{d\varphi^2}$ and each vertex corresponds to $v_n(\varphi)\equiv \frac{d^nV(\varphi)}{d\varphi^n}$, where $n$ is equal to the number of propagators at each vertex. The above Feynman rules for vertices can be found from the expansion of the potential $V(\varphi+\widehat \varphi)$ over the quantum field $\widehat\varphi$. As a result, the effective potential is constructed as a perturbative expansion over the coupling constant $g$
\begin{equation}
V_{eff}(\varphi,g)=g\sum_{n=0}^\infty (-g)^n V_n(\varphi).
\end{equation}

First of all, we compute the one-loop diagram corresponding to $V_1$. To regularize the UV divergence, we use dimensional regularization and take  $D=4-2 \epsilon$. Then one has 
\begin{equation}
V_1=  \frac{1}{16\pi^2}\frac 14\frac{v_2^2}{\epsilon} \left(\frac{\mu^2}{m^2}\right)^\epsilon \to \frac{1}{16\pi^2}\frac{v_2^2}{4}\left(\frac 1\epsilon -\log \frac{m^2}{\mu^2}\right), \ \ \  m^2=gv_2(\varphi),
\end{equation}
where $\mu$ is the transmutation parameter.
The divergent part $\sim 1/\epsilon$ is usually absorbed into the corresponding counterterm which in renormalizable theories leads to renormalization of the coupling. In the non-renormalizable case, this counterterm  does not repeat the original Lagrangian and thus leads to the appearance of a new structure. This is then continued in the subsequent orders of perturbation theory so that one has an infinite series of new terms. 
The arising situation may cause questions and seem to be unacceptable for rigorous quantum field theory followers. We discussed our views on this topic in our papers~\cite{Kazakov:2022pkc,Kazakov:2023tii} , where the formalism was introduced. Here we repeat the main arguments for completeness of presentation.\\
 Indeed, in the non-renormalizable  case, one has infinite arbitrariness in the choice of counterterms, which is not reduced to the renormalization of a finite number of couplings.  This arbitrariness appears in each order of perturbation theory where it can be fixed  by normalization of  a finite number of terms, however,  each new order brings still new couplings and the total number is infinite. There is no known procedure how one can fix this arbitrariness and we do not solve this problem. Let us stress  that we are not following the so-called effective field theory approach where this problem is not even touched and one simply takes the higher order operators of a give dimension and considers the energies below the scale of these operators  $M$ so that their contribution is suppressed be $E^2/M^2$.  \\
  Instead we assume that the problem of arbitrariness of subtractions is solved somehow and we notice that the leading terms of perturbative expansion (the leading logs) do nor depend on the subtraction scheme just like it takes place in renormalizable theories. This allows us to sum up the infinite series of the leading logs (LL) which has a universal behaviour and does not depend on the troubling infinite arbitrariness. Still we have some remnant of this arbitrariness, namely the LL have the form of $\log gv_2^2/\mu^2$, where $\mu$ is the unknown finite renormalization scale. This scale is the effect of dimensional transmutation and appears equally well in renormalizable theories. The variation of $\mu$ does not influence the LL approximation, this is the effect of the subleading order. So at the LL level one may claim the independence on $\mu$, though the explicit dependence is present. In renormalizable theories this explicit dependence is compensated up to a given order by implicit dependence of the coupling $g$ on $\mu$ via the RG equations.  Here it is not that simple, one has to vary an infinite sequence of couplings (finite in a given order of PT). As we have already mentioned, this problem is not yet solved. \\
    For this reason, in what follows we keep $\mu$ as free parameter, it may be fixed by some requirement. Then, we drop the the singular divergent terms and take care of the log dependence only, namely we  sum up the LL terms $g^n\log^n gv_2^2/\mu^2$ in all orders of PT. 

Note now that the coefficient of the singular term $\sim 1/\epsilon$ coincides with the coefficient of the logarithm $\sim \log(gv_2(\varphi))$ giving the contribution to the effective potential. This is true in all orders of PT. Hence, to calculate the leading log terms   $\sim\log^n \frac{m^2}{\mu^2}$, one has to calculate the leading poles $1/\epsilon^n$  and, replacing the pole by the log, get the leading contribution to the effective potential.  In its turn, to find the leading poles, one can use  the BPHZ $\cal{R}$-operation ~\cite{BogoliubovBook,Zimmermann,Hepp}  and the Bogoliubov-Parasiuk theorem \cite{BP} which states that after subtracting the subdivergences the singular part is local in coordinate space or has a polynomial structure in momentum space. This allows one to get recurrent relations for the subsequent leading divergences.
Denoting the singular part of the effective potential (the leading pole term $\sim 1/\epsilon^n$) in the n-th order of perturbation theory by $\Delta V_n$, we can write the following recurrence equation~\cite{Kazakov:2023tii}:
\begin{equation}
n \Delta V_n=  \frac 14 \sum_{k=0}^{n-1} D_2  \Delta V_k D_2  \Delta V_{n-1-k}, \ \ \, n\geq1, \ \   \Delta V_0=V_0 ,  \label{rec2}
\end{equation}
where $D_2\equiv\partial^2/\partial \varphi^2$ is the second derivative with respect to the field $\varphi$. 

This recurrent relation can be converted into the differential equation for the sum
\begin{equation}
\Sigma(z,\varphi)=\sum_{n=0}^{\infty} (-z)^n\Delta V_n(\varphi),\label{sum}
\end{equation}
where $z=g/\epsilon$. This is just the RG equation for the  function $\Sigma(z,\varphi)$
\begin{equation}
 \frac{\partial \Sigma}{\partial z}=-\frac 14 \left(\frac{\partial^2 \Sigma}{\partial \varphi^2}  \right)^2, \ \ \Sigma(0,\varphi)=V_0(\varphi). \label{RG}
\end{equation}
This partial differential equation generalises the well-known RG-equation in renormalizable scalar theory for the non-renormalizable case \cite{Kazakov:2023tii}. It is universal and does not depend on the arbitrariness of subtracting the UV divergences, as it was mentioned earlier. Unfortunately, except for the case of the renormalizable $\varphi^4$ potential, one cannot be solved analytically and one has to resort to numerical solution. 

To obtain an effective potential, one only needs to replace in the obtained solution the argument $z$ by the corresponding logarithm:
\begin{equation}
V_{eff}(g,\varphi)=g\Sigma(z,\varphi)|_{z\to -\frac{g}{16\pi^2}\log{gv_2/\mu^2}}.\label{effpot}
\end{equation}

The effective potential obtained this way contains the leading log contributions from all orders of perturbation theory. The full solution usually has smoother behaviour than the one-loop approximation and can include both poles and discontinuities and restore the original symmetry of the classical potential~\cite{Kazakov:2022pkc}. 

We now apply the formalism described above to study the global properties of effective potentials for the cosmological models of chaotic inflation.

\section{Cosmological constant from quantum corrections} 

Consider the following general action of the single-field inflationary cosmology:
\begin{equation}
    S = \int d^4x\sqrt{-{\textsl{g}}} \left[ \frac{M_{Pl}^2}{2}R({\textsl{g}}) + \frac{1}{2}{\textsl{g}}^{\mu\nu}\partial_\mu \phi \partial_\nu \phi - V(\phi)  \right] \label{STaction},
\end{equation}
where $\phi$ is the inflaton field, ${\textsl{g}}_{\mu\nu}$ is the space-time metric, $R({\textsl{g}})$ is the scalar curvature, ${\textsl{g}}=\det({\textsl{g}}_{\mu\nu})$ and $V(\phi)$ is the inflaton potential. In~\cite{Kallosh:2013hoa} it was shown that there exists a wide class of so-called $\alpha$-attractor models with hyperbolic geometry of the following form
\begin{equation}
    S = \int d^4x\sqrt{-{\textsl{g}}} \left[ \frac{M_{Pl}^2}{2}R({\textsl{g}}) + \frac{1}{2}\frac{\partial_\mu \phi \partial^\mu \phi}{1-\frac{\phi^2}{6\alpha}} - V(\phi)  \right] \label{STaction2},
\end{equation}
where $\alpha$ is a free parameter, that is in excellent agreement with the observational data.
Such geometries are natural in extended supergravity, which is one of the justifications of this action \cite{Kallosh:2013hoa}. 

One can get the canonical form of the kinetic term in eq.(\ref{STaction2}) by solving the equation $\partial \phi/\left(1-\frac{\phi^2}{1-6\alpha}\right)=\partial\varphi$ with the solution $\phi=\sqrt{6\alpha}\tanh\left(\frac{\varphi}{\sqrt{6 \alpha}}\right)$. Then the action (\ref{STaction2}) can be written in terms of the canonical field $\varphi$ with the canonical kinetic term as
\begin{equation}
    S = \int d^4x\sqrt{-{\textsl{g}}} \left[ \frac{M_{Pl}^2}{2}R({\textsl{g}}) + \frac{1}{2}\partial_\mu \varphi \partial^\mu \varphi - V\left(\sqrt{6\alpha}\tanh\left(\frac{\varphi}{\sqrt{6 \alpha}}\right)\right)  \right]. \label{STaction3}
\end{equation}
Therefore the previously mentioned potentials $ V_T= \tanh^{n}\left(\frac{\varphi}{\sqrt{6 \alpha} M}\right)$ correspond to the power like potentials for the field $\phi$, namely $\phi^n$.

As stated earlier, quantum corrections in general can significantly affect the behaviour of the potential, in particular, they can  "uplift" the minimum introducing the cosmological constant naturally rather than by hand \cite{Kazakov:2023tii}. This turns out to be possible due to the fact that quantum corrections cause the appearance of an additional parameter of dimensional transmutation $\mu$. In the non-renormalisable theory, the scale $\mu$ cannot be eliminated by the standard renormalization condition, unlike renormalisable theory \cite{Buchbinder:1984vys}, therefore, we can try to consider this parameter $\mu$ as inherent in the theory itself and as a free additional parameter. This is an important difference between such a theory and a renormalisable one. Below we consider how this mechanism works in the case of $T$ and $E$ models of chaotic inflation introduced above.

\subsection{The $T$-model}

To start with, we derive the generalised RG-equation for  the general $T$-model  with the potential $ g V_T(\varphi)=g \tanh^{n}\left(\frac{\varphi}{\sqrt{6 \alpha} M_{Pl}}\right)$ for an arbitrary power $n$.
To transform the equation (\ref{RG}) into an equation for the chosen potential, it is convenient to rewrite the function $\Sigma$ in terms of the dimensionless variables $x=z/M_{Pl}^4$ and $y=\tanh^{n}(\varphi/\sqrt{6 \alpha}M_{Pl})$. This substitution becomes obvious if one considers the loop decomposition of the $\Sigma$ function, which can be represented as polynomials over $\tanh\varphi$. Then one has
\begin{equation}
\Sigma(z/M_{Pl}^4, \tanh^n(\varphi/\sqrt{6 \alpha}M_{Pl}))\equiv S(x,y).
\end{equation}
This is somewhat similar to the arbitrary function $F(\tanh(\frac{\varphi}{\sqrt{6 \alpha}M_{Pl}}))$ of the $\alpha$-attractor-type models \cite{Kallosh2013}; however, this function is now constrained by the RG-equation.

Changing the variables in the original equation \eqref{RG}, the generalised RG-equation for the function $S(x,y)$ can be written as 
\begin{equation}
    S_x= -\frac{n^2 y^{2-\frac{4}{n}} \left(y^{2/n}-1\right)^2}{144 \alpha ^2} \left(\left(y^{2/n}+n \left(y^{2/n}-1\right)+1\right) S_y+n y \left(y^{2/n}-1\right) S_{yy}\right)^2.\label{12}
\end{equation}
The initial conditions are chosen from the requirement that the expansion of the effective potential starts from the classical potential in the limit of $\varphi \to \infty$, i.e. $\tanh(\varphi/\sqrt{6 \alpha M_{Pl}})\to 1$ the potential reaches the plateau with a zero derivative, namely
\begin{equation}
S(0,y)=y, ~ S(x,1)=1, ~ S_{y}(x,1)=0.
\end{equation}
Consider now the case when $n=2,4$, which we denote as $T^2$- and $T^4$- models. The RG equations for theses cases were first derived in \cite{Kazakov:2023tii}. One gets for
$n=2$
\begin{equation}
    S_x=-\frac{(y-1)^2 \left((3 y-1)S_y+2 (y-1) y S_{yy}\right)^2}{36\alpha^2} \label{14},
\end{equation}
and for $n=4$
\begin{equation}
   S_x= -\frac{y \left(4 y \left(\sqrt{y}-1\right)^2 S_{yy}+\left(5 y-8 \sqrt{y}+3\right) S_y\right)^2}{9\alpha^2}\label{15}.
\end{equation}

Now one has to solve these partial differential equations numerically for the function of two variables and then to choose on the surface $(x,y)$ the line corresponding to $x= -\frac{g}{16\pi^2}\log{gv_2(\varphi)/\mu^2}$. This line has an additional parameter, the scale $\mu$. In the renormalizable case, the explicit dependence on this scale is compensated by implicit dependence of the coupling $g$. However, in the non-renormalizable case, $\mu$ remains a free parameter. For various values of $\mu$ one can have different behaviour of the potential. In what follows we will choose the value of $\mu$ to satisfy the slow-roll inflation conditions.

The behaviour of the effective potential for the $T^2$-model in comparison with the one-loop case is shown in Figure \ref{fig:oneloop}. 
\begin{figure}[ht]
\begin{center}
\includegraphics[scale=0.6]{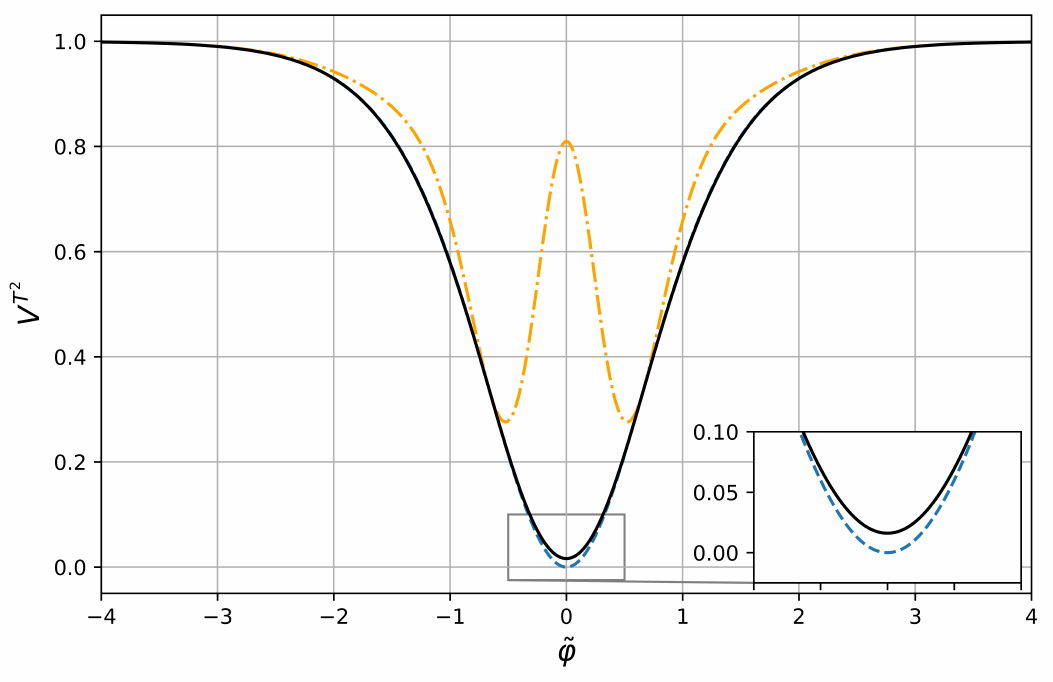}
\caption{Comparison of the classical $T^2$-model potential (blue dashed line), the one-loop correction (orange dashed line), and the RG effective potential (black solid line) for illustration at $g \sim 1, ~ \mu <M_{Pl}$, $\tilde{\varphi} =\varphi/(\sqrt{6}M_{Pl})$} \label{fig:oneloop}
\end{center}
\end{figure}
One can see how the one-loop quantum correction leads to a change of the vacuum state  of the potential and to the appearance of additional minima (this is referred to as the Coleman-Weinberg mechanism \cite{CW}) when choosing some values of the parameter $\mu$. However the sum of all leading quantum corrections smoothes out the behaviour of the potential and restores the original symmetry. Note that the plateau-like behaviour of the potential is not changed  whichever approximation of the effective potential we consider \cite{Kallosh:2016gqp}. We have studied various Hubble flow parameters and found out that quantum corrections do not change the slow-roll behaviour of the potential for a wide range of parameters $\alpha,g$ and $\mu$ \cite{Kazakov:2023tii}.

We show also in Fig.\ref{mu} a more detailed view of the full effective potential for various values of $\mu$.
Our aim is to get the uplifting of the potential to obtain non-zero value of energy at the minimum.
\begin{figure}[h]
\begin{center}
\includegraphics[width=11cm]{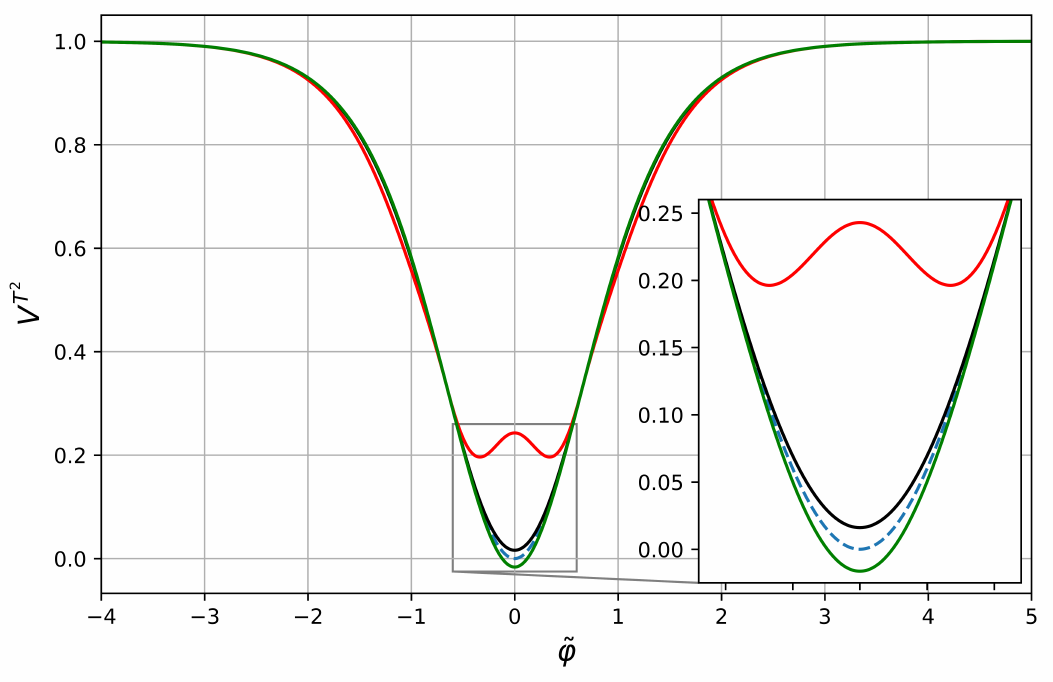}
\caption{$T^2$-model potential: variation of $\mu$. The classical potential (blue dashed line), solid lines denotes all-loop effective potential  $\mu < M_{Pl}$ (black line), $\mu \ll M_{Pl}$ (red line), and $\mu > M_{Pl}$ (green line). The scales  are chosen for illustrative purposes as follows $\tilde{\varphi} =\varphi/(\sqrt{6}M_{Pl})$ and $g=2$, $\alpha=1$}
\label{mu}
\end{center}
\end{figure}

As can be seen from Fig.\ref{mu}, there are mainly two viable possibilities to get a positive value of $\Lambda$. First, when the effective potential does not have additional minima, but has only some uplift. And the second possibility is when the potential, in addition to the uplift, has additional minima, and the uplift occurs in these minima. The realisation of the first and second variants depends only on the value of $\mu$.

In the case when the full result does not differ qualitatively from the one-loop correction, it is convenient to use explicit formulas for the potential to study its properties. The one-loop effective potential has the form
\begin{equation}
V_{eff}=V_0+\frac{g^2}{16\pi^2}\frac{v_2^2}{4}\log \frac{gv_2(\varphi)}{\mu^2}, 
\end{equation}
where $v_2(\varphi)$ is the second derivative of the classical potential $V_0$.

The non-zero value of the potential at the minimum may be associated with the cosmological constant given by  the relation
\begin{equation}
    \Lambda=\left.\left[\frac{g^2}{16\pi^2} \frac{v_2^2}{4} \log\left(\frac{g v_2}{\mu^2}\right)\right]\right\vert_{\varphi=\varphi_{vac}}\label{16}.
\end{equation}
Applying this condition to the $T^2$-model, one can get  the expression for the parameter $\mu$ 
\begin{equation}
    \mu^2=\frac{g}{3 \alpha  M_{Pl}^2}e^{-576 \pi ^2\frac{ \alpha ^2 }{g^2} \Lambda  M_{Pl}^4}.
\end{equation}
Choosing the following set of parameters corresponding to modern observational data $g=10^{-10}M_{Pl}^4$, $M_{Pl}=(8\pi G)^{-\frac{1}{2}}$, $\alpha=1$, and $\Lambda \sim 10^{-120} M_{Pl}^4$, we obtain that $\mu \approx 10^{-6}M_{Pl}$, which is  reasonably close to the inflaton mass.
One can invert this argument and take the value of $\mu$ of the order of the inflaton mass and get a reasonably small value of  the cosmological constant.

When $\Lambda$ is as small as it follows from experimental data, the value of $\mu$ practically does not depend on the precise value of  $\Lambda$. The variation of  $\Lambda$  corresponding to the observed data changes the value of $\mu$ only by a few  decimal points. 
However, if we start to increase $\Lambda$, we find that starting from  about $10^{-24}M_{Pl}^4$, the $\mu$ parameter starts to depend very strongly on  $\Lambda$, and $\mu$ quickly tends to zero at  $\Lambda\approx 10^{-20}M_{Pl}^4$.

At the same time, it is possible to  calculate the point where $\Lambda$ is exactly zero, i.e. the point where it changes the sign.  This point is found from the condition of changing the sign of the logarithm in \eqref{16}, namely, when the argument passes through unity, i.e. $\left.\frac{g v_2}{\mu_c^2}\right\vert_{\varphi=\varphi_{vac}}=1$ and we have the critical point 
 \begin{equation}
 \mu_{c}=\frac{10^{-5}}{\sqrt{3 \alpha}}M_{Pl}.
 \end{equation}
When $\mu>\mu_c$, $\Lambda$ is negative, when $\mu<\mu_c$, $\Lambda$ is positive. This scale is comparable with the scale of the inflaton mass. The critical value of $\mu$ determines the geometry in the action \eqref{STaction}, namely: de Sitter space (dS) when $\mu>\mu_c$, Minkowski space when $\mu=\mu_c$ and anti-de Sitter space (AdS) when $\mu<\mu_c$.  

A qualitatively similar situation takes place when one takes not only the one-loop contribution to the effective potential but also the full
solution that includes the leading logs from all loops. In this scenario, the effective potential has the only minimum at $\varphi_{vac}=0$ and no additional ones. At the minimum one has a tiny uplift for small values of $\mu$. For large values of $\mu$ one can have additional minima, but it is difficult to get the desired small values of $\Lambda$.

Regarding the potential $T^4$ and higher powers, they do not have the same uplift of the potential as in the case of the $T^2$-model, which can be seen from the numerical solution 
shown in Fig.\ref{T4}. This is because $v_2(0)=0$. Moreover, at additional minima the value of $\Lambda$ happens to be negative.  So these potentials are not suitable for describing the cosmological constant.

\begin{center}
\begin{figure}
\centering
\includegraphics[width=12cm]{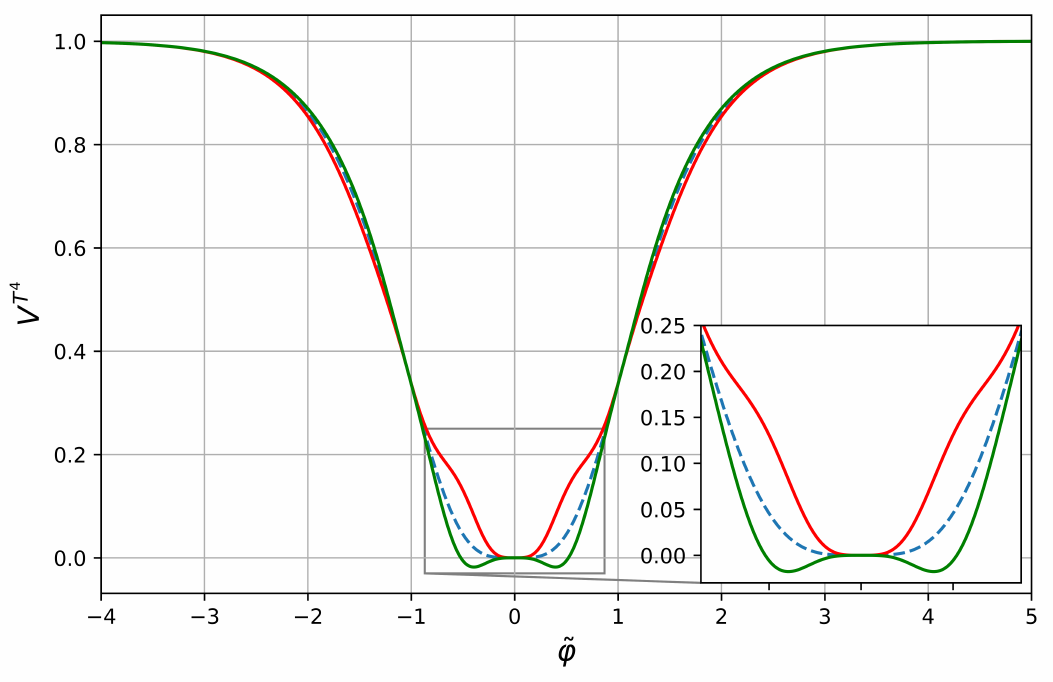}
\caption{Classical $T^4$-model potential (blue dashed line), solid lines denote the all-loop effective potential: red one at $\mu^2 \ll M_{Pl}^2 $ and  green one  at $\mu^2 \gg M_{Pl}^2 $. The scales  are chosen for illustrative purposes as follows $\tilde{\varphi} =\varphi/(\sqrt{6}M_{Pl})$ and $g=1$, $\alpha=1$}
\label{T4}
\end{figure}
\end{center}

\subsection{The $E$-model}
Here we repeat the same analysis of the potentials for the $E$-model (\ref{potE}).
This potential is interesting because it is asymmetric and plateau-like behaviour is present only for positive values of the field, while for negative values the potential grows exponentially fast to infinity. From qualitative considerations it is not quite clear whether such asymptotic behaviour of the potential will remain intact after taking into account quantum corrections.

As before, to simplify equation (\ref{RG}), we rewrite  the function $\Sigma$  in the dimensionless variables $x=z/M_{Pl}^4$ and $y=(1-e^{\sqrt{\frac{2}{3\alpha}}\frac{\varphi}{M_{Pl}}})^{n}$. Then one has
\begin{equation}
\Sigma\left(z/M_{Pl}^4, (1-e^{\sqrt{\frac{2}{3\alpha}}\frac{\varphi}{M_{Pl}}})^{n}\right)\equiv S(x,y).
\end{equation}
Changing the variables in the original equation \eqref{RG}, the generalised RG-equation for these potentials can be written as 
\begin{equation}
    S_x=-\frac{n^2}{9 \alpha ^2} y^{2-\frac{4}{n}} \left(y^{1/n}-1\right)^2 \left(\left(n \left(y^{1/n}-1\right)+1\right) S_y+n y \left(y^{1/n}-1\right) S_{yy}\right)^2.  \label{21}
\end{equation}
Taking the cases $n=2,4$ we obtain for $n=2$:
 \begin{equation}
     S_x=-\frac{4}{9 \alpha^2} \left(2 y \left(\sqrt{y}-1\right)^2 S_{yy}+\left(2 y-3 \sqrt{y}+1\right) S_y\right)^2 \label {22},
 \end{equation}
and for $n=4$:
 \begin{equation}
        S_x=-\frac{16}{9 \alpha^2}\left(\sqrt[4]{y}-1\right)^2 y \left(\left(4 \sqrt[4]{y}-3\right) S_y+4 y \left(\sqrt[4]{y}-1\right) S_{yy}\right)^2\label {23},
 \end{equation}
with the initial conditions given by 
 \begin{equation}
    S(0,y)=y, S(x,1)=1, S_{y}(x,1)=0.
\end{equation}
As before, the cases  of $n=4$ and higher are unsuitable for the description of the cosmological constant, because there is no corresponding uplift due to $v_2(0)=0$.  

From Fig.\ref{deE2} it is easy to see that in the case at small values of $\mu$ the behaviour of the potential repeats the classical case except for a small modification of the slow rolling inflation region. 
One can notice some similarity between the behaviour of the potential for the $E^2$-model and the potential given in  \cite{Scalisi:2015qga} with small changes of $g,\alpha$ and $\mu \ll M_{Pl}$.
\begin{figure}[ht]
\centering
\includegraphics[width = 11.cm]{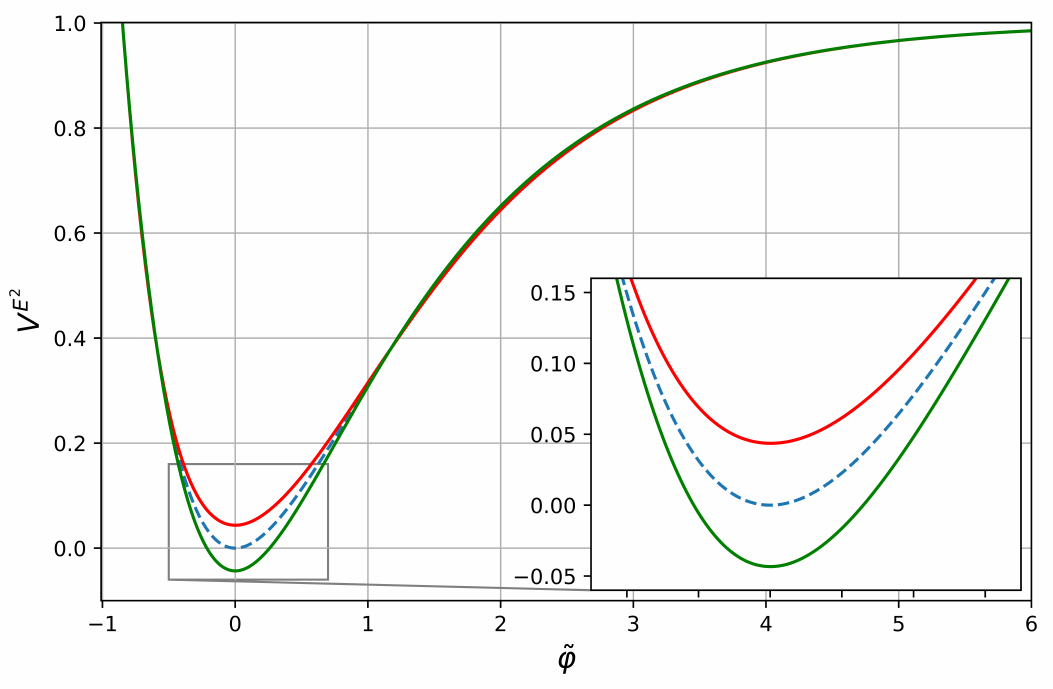}
\caption{Classical $E^2$- model potential (red dashed line) and its RG-corrected versions: the red line is for $\mu^2 < M_{Pl}^2$ and the green line is for $\mu^2>M_{Pl}^2$. The scales  are chosen for illustrative purposes as follows $\tilde{\varphi} =\varphi/(\sqrt{6}M_{Pl})$ and $g=1$, $\alpha=1$}
\label{deE2}
\end{figure}

As in the previous case, we can estimate the  value of $\mu^2$ from the one-loop expression for the $E^2$-model (Fig.\ref{deE2}) and get
\begin{equation}
    \mu^2=\frac{4 g }{3 \alpha  M_{Pl}^2} e^{-36 \pi ^2\frac{ \alpha ^2}{g^2} \Lambda  M_{Pl}^4}\label {30},
\end{equation}
for $\alpha=1$ this gives  $\mu\approx10^{-5}M_{Pl}$ for small $\Lambda$. In this case the minimum of the potential is only slightly shift at small values of $\mu$. But from the numerical solution of equation (\ref{22}) one gets $\varphi_{vac}=0$. As in the previous case, no additional minima appear.  As before, one can calculate the critical point where $\Lambda$ changes its sign:
 \begin{equation}
 \mu_c=2\frac{10^{-5}}{\sqrt{3 \alpha}}M_{Pl}.
 \end{equation}
which is again close to the inflaton mass.

\subsection{The model of quintessence}

Consider at last an even simpler potential $T^{1}$ corresponding to the quintessence  model \cite{Akrami:2017cir}
\begin{equation}
    V=\gamma \sqrt{6\alpha}\left(\tanh\left(\frac{\varphi}{\sqrt{6 \alpha} M}\right)+1\right),\label {Tl}
\end{equation}
which  has the shape shown in Fig.\ref{deE1} (plus 1 is added to get the asymptotics as $\varphi \to -\infty$ equal to zero).
\begin{figure}[ht]
\centering
\includegraphics[width = 15.cm]{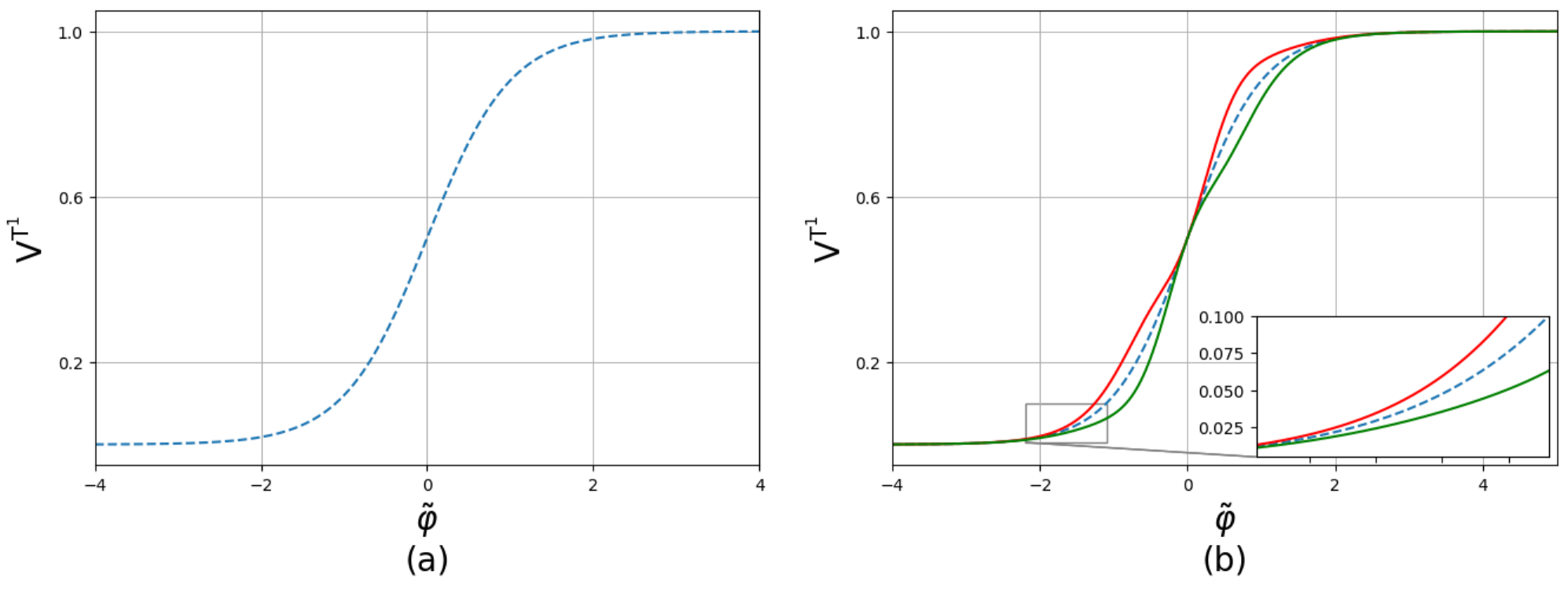}
\caption{The potential of the quintessence model: classical potential (a), potential with quantum corrections (b). The green line is for $\mu \gg M_{Pl}$ and the red line is for $\mu \ll M_{Pl}$. The scales  are chosen for illustrative purposes as follows $\tilde{\varphi} =\varphi/(\sqrt{6}M_{Pl})$ and $g=10$, $\alpha=1$}
\label{deE1}
\end{figure}

In the presence of the cosmological constant one should expect that $V\vert_{\varphi=-\infty}=\Lambda$~\cite{Akrami:2017cir}. We check whether this scenario is realized when taking into account quantum corrections. The numerical solution of eq.(\ref{12}) for the case
of $n=1$
\begin{equation}
S_x=-\frac{(y^2-1)^2}{144\alpha^2y}(2yS_y+(y^2-1)S_{yy})^2    
\end{equation}
is shown in Fig.\ref{deE1} (right). One can see that in this case, as 
in the other $\alpha$-attractor models, the asymptotics  of the classical potential is preserved when taking into account all quantum corrections\cite{Kazakov:2023tii}: it goes to 1 when $\varphi\to +\infty$ and to zero when $\varphi\to -\infty$ without any additional minima. The only influence of quantum corrections is a change in the shape of the potential in the middle while preserving the limit as $\varphi \to \pm \infty$.  This can be interpreted as an additional contribution of the effective field-dependent cosmological constant $\Lambda(\varphi)$ which decreases with time \cite{Akrami:2017cir} to the value of the dark energy determined by the classical potential (\ref{Tl}).

To find the contribution of quantum corrections to $\Lambda(\varphi)$,  one can again use the one-loop potential given by the same formula
\begin{equation}\Lambda(\varphi)=\frac{g^2}{16\pi^2} \frac{v_2^2}{4} \log\left(\frac{g v_2}{\mu^2}\right),
\end{equation}

In this way, it is possible to fit the parameter $\mu$ to get a modification of the original classical potential that might be required by, say, observational data and restrictions on quintessential inflation. Therefore, even such a simple $T^1$ model of quintessence, with taking into account quantum corrections, can be quite flexible for description of observational data.

\section{Conclusions}
In this paper, we have considered the influence of quantum loop corrections on the potentials of the $\alpha$-attractor models that are often used in inflationary cosmology. For this purpose, we derived the generalised RG equations  (\ref{12},\ref{14},\ref{15},\ref{21}-\ref{23}) that sum up the whole sequence of leading logarithmic contributions to the effective potential.  Solving these equations, one can take into account all quantum corrections  that usually smooth out the behaviour of the one-loop contribution and,  depending on the value  of the free scale parameter $\mu$,  either preserve the minimum of the potential at the origin or  create new minima but in both cases lift the potential, thus producing the cosmological constant.

We mainly focused our attention on the effect of uplifting the effective potential due to quantum corrections. This uplift, which appears in the one-loop approximation, remains intact even if all quantum corrections taken into account. We interpreted this uplift as the appearance of the cosmological constant $\Lambda$ for the $T^{2}$ and $E^{2}$ models and as an additional contribution to the dark energy for the $T^{1}$ model. The obtained value of the free parameter $\mu\approx10^{-5}M_{Pl}$, which was calculated in the one-loop approximation, lies in an interesting energy region being close to the inflaton mass. Numerical evaluation of the equations only slightly changes this value.  It is also important to note that the value of the parameter $\mu$ plays a crucial role in determining the geometry of space-time: flat, de Sitter space (dS) or anti-de Sitter space (AdS). The critical value of $\mu$ can be easily evaluated for the models under consideration.

Thus, we have found out that the cosmological constant $\Lambda$ may exist as a consequence of quantum corrections to the effective potential  with some value of the scale parameter $\mu$ even in non-renormalizable models of inflation. The value of  the cosmological constant $\Lambda$ allows  one  to fix the  parameter $\mu$ which is a free parameter in the non-renormalizable theory. One may hope that the mechanism of radiative corrections can shed light on the existence of Dark energy in the early Universe.

\section{Acknowledgements} The authors are grateful to
I.Buchbinder, S.Fedoruk, and A.Baushev  for valuable discussions.

\bibliographystyle{JHEP}
\bibliography{biblio.bib}
\end{document}